\documentclass[conference]{IEEEtran}
\IEEEoverridecommandlockouts
\usepackage{cite}
\usepackage{amsmath,amssymb,amsfonts}
\usepackage{graphicx}
\usepackage{algorithm}
\usepackage{algpseudocode}
\usepackage{textcomp}
\usepackage{url}
\usepackage{multirow}
\usepackage{tcolorbox}
\usepackage{booktabs}
\usepackage{tcolorbox}
\tcbuselibrary{skins}
\usepackage{multirow}
\usepackage{bbding}
\usepackage{utfsym}
\usepackage{ulem}
\usepackage{balance}
\usepackage{wasysym}
\usepackage{hyperref}
\usepackage{multirow}
\usepackage{makecell}
\usepackage{pifont}
\usepackage{fontawesome5} 

\usepackage[utf8]{inputenc}
\usepackage[T1]{fontenc}

\def\eg{\textit{e.g.,} }
\def\ie{\textit{i.e.,} }

\newtcolorbox{AcademicBox}[1][]{academicbox=#1}
\definecolor{SoftBlue}{RGB}{135, 206, 250}  
\definecolor{SoftOrange}{RGB}{255, 224, 178} 
\definecolor{SoftGreen}{RGB}{144, 238, 144}  
\definecolor{CorrectGreen}{RGB}{76, 175, 80} 
\definecolor{ErrorRed}{RGB}{211, 47, 47} 

\def\BibTeX{{\rm B\kern-.05em{\sc i\kern-.025em b}\kern-.08em
    T\kern-.1667em\lower.7ex\hbox{E}\kern-.125emX}}

\tcbset{
  colframe=black!10,
  colback=white,
  coltitle=black,
  fonttitle=\bfseries,
  boxrule=0.4pt,
  left=6pt,right=6pt,top=6pt,bottom=6pt,
  title style={left color=black!3,right color=black!1}
}

\AtBeginDocument{%
  \providecommand\BibTeX{{%
    Bib\TeX}}}
\newtcolorbox{boxK}{
    top=2pt,
    bottom=2pt,
    left=2pt,
    right=2pt,
    boxrule = 0pt,
    toprule = 0pt, 
    colback=gray!15,   
    colframe=white     
}

\tcbset{
  my box/.style={
    enhanced,
    colframe=#1!80,
    colback=#1!10,
    attach boxed title to top left={xshift=0.2cm, yshift=-0.2cm},
    boxed title style={
      colback=#1!80,
      coltext=white,
      outer arc=0pt,
      arc=0pt,
      top=0pt,
      bottom=0pt,
    },
  },
}

\newtcolorbox{Prompt}[1]{
  my box=black,
  title=#1,
  coltitle=white,
  boxrule=1.2pt,top=6pt,bottom=3.5pt,left=6pt,right=6pt
}

\begin{document}

\def\eg{\textit{e.g.,} }
\def\ie{\textit{i.e.,} }

\newcommand{\td}[1]{{\color{blue}{\textbf{#1}}}}
\newcommand{\ours}[1]{\textit{OntoAgent}}

\title{From Chat to Interview: Agentic Requirements Elicitation with an Experience Ontology} 


 \author{
 \IEEEauthorblockN{Dongming Jin$^{1,2}$, Zhi Jin*$^{1,2}$\thanks{*Corresponding authors}, Yaotian Yang$^{3}$, Linyu Li$^{1,2}$, Zheng Fang$^{1,2}$, Yuanpeng He$^{1,2}$, \\ Wenchun Jing$^{1,2}$, Xiaohong Chen$^{4}$}
    \IEEEauthorblockA{$^1$ Key Laboratory of High Confidence Software Technologies (Peking University), Ministry of Education, China}
    \IEEEauthorblockA{$^2$ School of Computer Science, Peking University, China}
    \IEEEauthorblockA{$^3$ Beijing Forestry University, China}
    \IEEEauthorblockA{$^4$ East China Normal University, China}
    \IEEEauthorblockA{\texttt{correspondence to: zhijin@pku.edu.cn}}
}

\maketitle

\begin{abstract}
Requirements elicitation interviews are crucial and time-consuming in requirements engineering, but heavily rely on the experience of requirements analysts. Although recent advancements in large language models (LLMs) have created new opportunities to automate this process, existing approaches rely solely on LLMs for free-form chat without taking into account the interview and development experience. That leads to the omission of implicit requirements and redundant questions. Practically, experienced analysts implicitly follow a structured cognitive framework when conducting requirements elicitation. 

Inspired by this observation, this paper proposes an interview agent named \ours{} for the elicitation of requirements guided by an experience ontology. \ding{182} \ours{} automatically analyzes domain-specific requirements descriptions to construct an experience ontology, which organizes requirements concerns into an ontology to support systematic and explainable interviews. \ding{183} During the interview, \ours{} first performs four operations (\ie ParseUser, ScoreOnto, ReRankOnto, GatePrune) guided by the ontology to identify the relevant requirement concerns. The selected concern is then combined with the current dialogue context to generate the elicitation question. To validate \ours{}, we conduct comprehensive quantitative experiments using the widely adopted website application domain. The results show that \ours{} significantly outperforms existing baselines in both elicitation effectiveness and questioning efficiency, achieving a 33\% improvement in IRE and a 21\% improvement in TKQR. Ablation studies further validate the contribution of each key design component. In addition, a qualitative user study demonstrates its practical advantages in real-world scenarios. We believe that \ours{} can also be extended to requirements interview tasks in other domain\footnote{To support reproducibility, we publicly release our code, dataset, and a lightweight tool at \url{https://anonymous.4open.science/r/TypoAgent-RE2026}.}. 
\end{abstract}

\begin{IEEEkeywords}
Requirements Elicitation Interview, Requirements Agent, Large Language Model, Ontology Engineering
\end{IEEEkeywords}

\section{Introduction}






Requirements elicitation is one of the core activities in requirements engineering, laying the foundation for accurately capturing stakeholders’ needs~\cite{goguen1993techniques}. Among various methodologies, interviews remain the most widely adopted elicitation technique~\cite{wiegers2013software}~\cite{shen2025requirements}. However, effective interviews heavily rely on experienced requirements analysts. This not only requires substantial time investment but also incurs significant training and labor costs~\cite{shen2025requirements}~\cite{meth2013state}. Moreover, interviews are inherently vulnerable to human bias and communication misunderstandings, which may lead to incomplete or ambiguous requirements~\cite{KornGV25}. These challenges become even more pronounced in the era of rapidly evolving AI-assisted software development. Therefore, \textbf{automating or intelligently supporting requirements elicitation interviews} has become an urgent and important research problem in requirements engineering.

\begin{figure}
    \centering
    \includegraphics[width=0.99\linewidth]{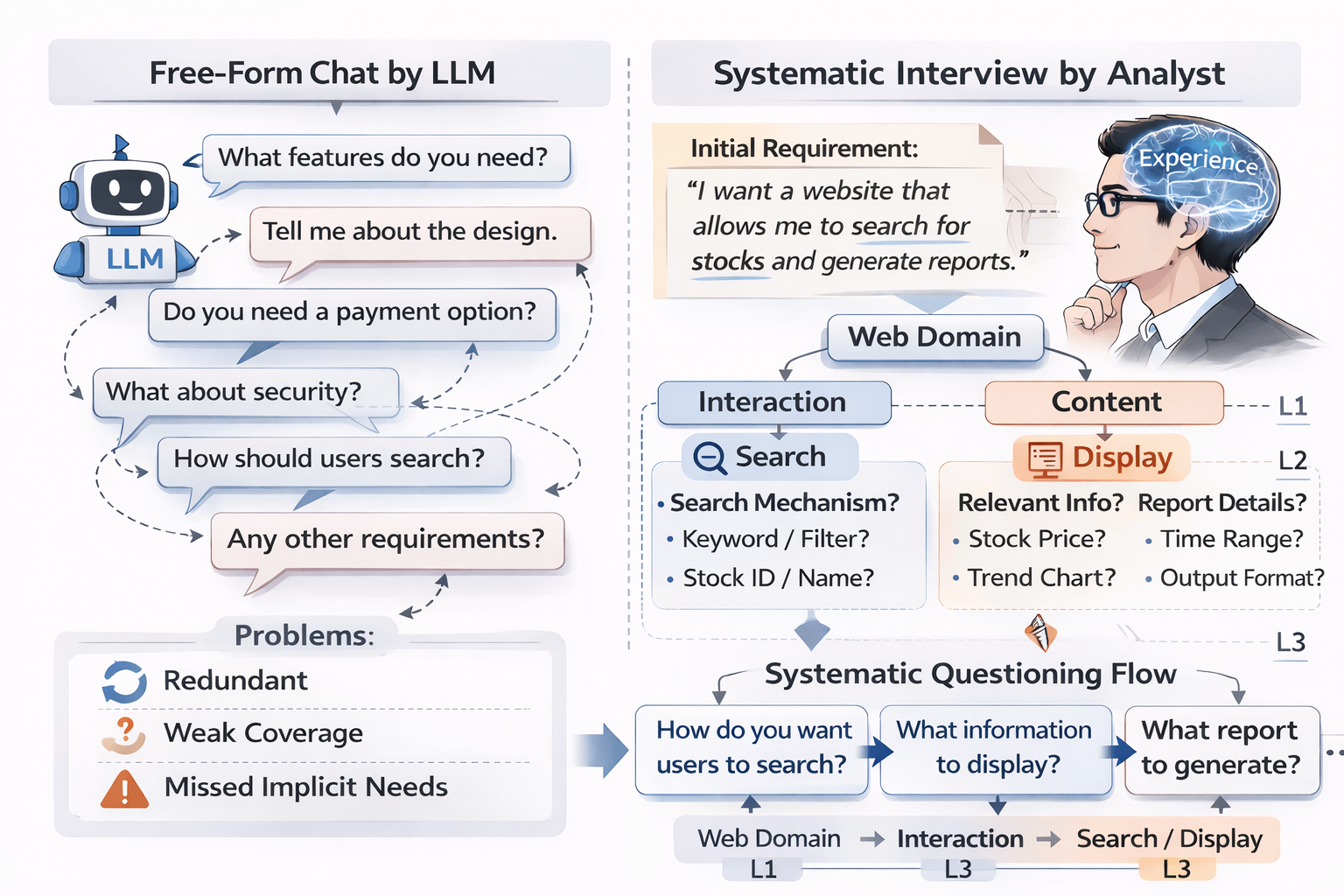}
    \caption{\textbf{Motivation of this work.} Left: Free-form LLM chat produces ad-hoc questions, resulting in redundancy and incomplete coverage of implicit requirements; Right: Experienced analysts implicitly follow a structured interview experience to systematically explore requirement dimensions}
    \label{fig:teaser}
\end{figure}

With the rapid advancement of large language models (LLMs) in dialogue generation~\cite{srinivas2025substance}~\cite{zhao2025personalens} and role-playing capabilities~\cite{jin2025iredev}~\cite{ataei2025elicitron}, researchers have begun to explore using LLMs to automate requirements elicitation interviews~\cite{KornGV25}~\cite{jin2026reqelicitgym}, i.e., as shown in the left part of Figure~\ref{fig:teaser}, an LLM is prompted to act as an interviewer and rely on its generative ability to ask free-form questions. In this paper, we refer to this approach as the free-form elicitation approach. Although it offers a straightforward solution, an empirical study~\cite{jin2026reqelicitgym} reveals that it suffers from two major limitations, \ie struggle to uncover implicit requirements and generate generic questions in early-stage dialogues.  In addition, the generated interview questions lack interpretability and controllability, making it difficult to incorporate effective human-in-the-loop supervision. These limitations negatively affect the effectiveness and efficiency of requirements elicitation. 


In contrast to the free-form elicitation approach, experienced requirements analysts implicitly follow a structured cognitive framework that is formed from long-term interview experience. This framework organizes requirement concerns into a hierarchical structure, thereby guiding systematic questioning to explore requirements. We refer to such structured knowledge as the experience ontology. The right panel of Figure~\ref{fig:teaser} illustrates the process of systematic interviews by experienced analysts. When facing an underspecified initial requirement such as \textit{``I want a website that allows users to search stocks and generate reports''}, experienced analysts first reason from high-level aspects of a Web application (e.g., \texttt{Interaction} and \texttt{Content}). Then they narrow the interview to specific functional dimensions (\eg \texttt{Search} and \texttt{Display}), and finally identify missing implementation slots within each dimension (\eg search mechanisms and report formats). Through this hierarchical expansion, the interview process forms a coherent and progressive questioning flow, improving coverage while reducing redundancy.

Inspired by the above structured interviewing process, we propose \ours{}, an interview agent for requirements elicitation guided by an experience ontology. Unlike the free-form elicitation, \ours{} explicitly models the interview experience as an ontology. This ontology serves as a structured questioning space to guide dynamic decision-making and question generation during interviews by decoupling \texttt{what to ask} from \texttt{how to ask}. The ontology determines which requirement slots should be explored, while the LLM generates natural language questions conditioned on dialogue context. Specifically, \ours{} consists of two core stages:

\begin{itemize}
    \item \textbf{Experience Ontology Induction.} Given a collection of domain-specific requirements texts, \ours{} automatically analyzes them to construct an experience ontology. The ontology is designed as a three-level hierarchical tree consisting of aspects, dimensions, and slots. 
    \item \textbf{Ontology-Guided Interviewing.} \ours{} conducts four key operations (\ie ParseUser, ScoreOnto, ReRankOnto, and GatePrune) over the experience ontology to dynamically prioritize and select the most relevant yet insufficiently explored requirement slots. According to the selected slots and the current dialogue context, it generates context-aware and targeted elicitation questions.
\end{itemize}
    
We conduct extensive experiments to evaluate \ours{}. (1) we evaluate \ours{} on 101 website requirements elicitation scenarios in ReqElicitGym~\cite{jin2026reqelicitgym}. We employ two complementary metrics, \ie Implicit Requirements Elicitation Ratio (IRE)~\cite{jin2026reqelicitgym} and Turn-discounted Key Question Rate (TKQR)~\cite{fangzcs_cognicode_2026}. Results show that \ours{} significantly outperforms previous baselines. \ours{} improves IRE by up to 33\%, demonstrating enhanced effectiveness in uncovering implicit requirements. \ours{} also improves TKQR by 21\%, indicating higher questioning efficiency and quality. (2) We perform ablation studies by incrementally adding four key design modules (\ie Experience Ontology, ScoreOnto, ReRankOnto, and GatePrune) to a base LLM. Results prove the independent contributions of each component. (3) We evaluate the sensitivity by switching six different base LLMs of \ours{}, showing strong generalization capability. (4) We conduct a human evaluation to evaluate the generated interview process in
three aspects, including elicitation effectiveness, efficiency, and adaptability. Results show that \ours{} outperforms baselines in all three aspects. (5) We conduct a case study to qualitatively compare the interview processes of \ours{} and baseline approaches. Results show \ours{} achieves more structured elicitation interviews in practical scenarios.

We summarize our contributions in this paper as follows. 

\begin{itemize}
        \item We propose \ours{}, an interview agent for requirements elicitation. It explicitly models the interview experience using ontology and integrates the experience ontology with LLMs to facilitate the generation of interview questions.
    \item We introduce and develop the interview experience ontology, a three-level hierarchical tree consisting of requirements aspects, dimensions, and slots, which serves as a structured and interpretable questioning space.
    \item We design an ontology-guided interviewing mechanism with four decision-making operations to dynamically select requirement slots and conduct the generation of the context-aware elicitation questions.
    \item We conduct extensive experiments on 101 requirements elicitation scenarios. Qualitative and quantitative analyses show the effectiveness and practical applicability of our \ours{}.
\end{itemize}

\section{background and related works}


Requirements elicitation is widely recognized as a pivotal yet challenging phase in requirements engineering, as inadequate understanding of stakeholder needs remains a major cause of project failure~\cite{jones2004software}. Among various elicitation approaches, interviews are one of the most traditional and commonly used techniques~\cite{wiegers2013software}~\cite{ferrari2016ambiguity}. It enables analysts to directly engage with stakeholders and ask questions to understand their needs, preferences, and expectations about a product or service~\cite{hickey2003elicitation}~\cite{palomares2021state}. During interviews, interviewers can ask questions to elicit stakeholder preferences, resolve possible ambiguities around multifaceted and conflicting viewpoints~\cite{zowghi2005requirements}, and gain insight into tacit knowledge the interviewees might harbor~\cite{ferrari2016ambiguity}, ultimately obtaining a list of requirements~\cite{ferrari2022requirements}. Despite the variety of interview formats, conducting interviews involves numerous challenges. In particular, stakeholder preferences are often tacit knowledge that requires significant effort to extract~\cite{ferrari2016ambiguity}. The existing literature has described various criteria in the appropriate conduct of interviews. The introduction to them is as follows. 

Before the emergence of LLMs, research primarily focused on supporting human analysts, including criteria for the appropriate opening and closing of interviews~\cite{bano2018learning}, the right atmosphere and flow~\cite{donati2017common}, the question framing~\cite{zaremba2021towards}, the question content~\cite{bano2018learning}, and the avoidance of common mistakes~\cite{han2021designing}. For example, Zaremba et al.~\cite{zaremba2021towards} proposed a set of systematic questions to improve interview preparation. Other studies investigated pedagogical strategies to enhance interview skills of humans, such as learning from common elicitation mistakes~\cite{bano2019teaching} and employing role-based training methods including role-playing, peer review, and self-assessment~\cite{ferrari2019learning,ferrari2020sapeer}. Additionally, Debnath et al.~\cite{debnath2022annoterei} developed AnnoteREI to facilitate transcription and annotation of interview data. Thus, these studies primarily aim to assist or enhance human-conducted interviews. In contrast, our work focuses on automating the requirements elicitation interview process with LLMs.

With the rapid advancement of LLMs, recent research has begun exploring intelligent support for requirements elicitation interviews. One line of work investigates using LLMs to generate interview scripts or question templates~\cite{gorer2023generating}. Another line leverages LLMs to extract or synthesize requirements directly from dialogue transcripts, aiming to reduce the manual effort required in requirements consolidation~\cite{almeida2025elicitation}. These studies demonstrate the strong capabilities of LLMs in supporting the preparation and recording. More recently, researchers have explored conducting interview directly using LLMs. KornGV et al.~\cite{KornGV25} investigated the feasibility of LLM-driven conversational elicitation and introduced two prompt-based interview chatbots named LLMREI-long and LLMREI-short. Similarly, Shen et al.~\cite{shen2025requirements} investigate LLM-based follow-up question generation for requirements elicitation and propose mistake-guided prompting to improve question quality. Additionally, Jin et al.~\cite{jin2026reqelicitgym} proposed an evaluation environment named ReqElicitGym for assessing interview competence in multi-turn dialogue settings, and conducted an empirical study on six mainstream LLMs. The results reveal that relying solely on free-form LLM chat often struggles to systematically uncover implicit requirements and generate generic questions in early-stage dialogues. Our work follows this direction, attempting to alleviate the aforementioned limitations and improve the effectiveness and interpretability of automated requirements interviews by explicitly integrating the structured interview experience with LLM-based question generation. 


\section{Approach}
This section presents \ours{}, the interview agent for requirements elicitation . We describe the overview of \ours{} approach in the first subsection and describe the details in the following subsections.

\subsection{Overview}
\begin{figure*}
    \centering
    \includegraphics[width=0.9\linewidth]{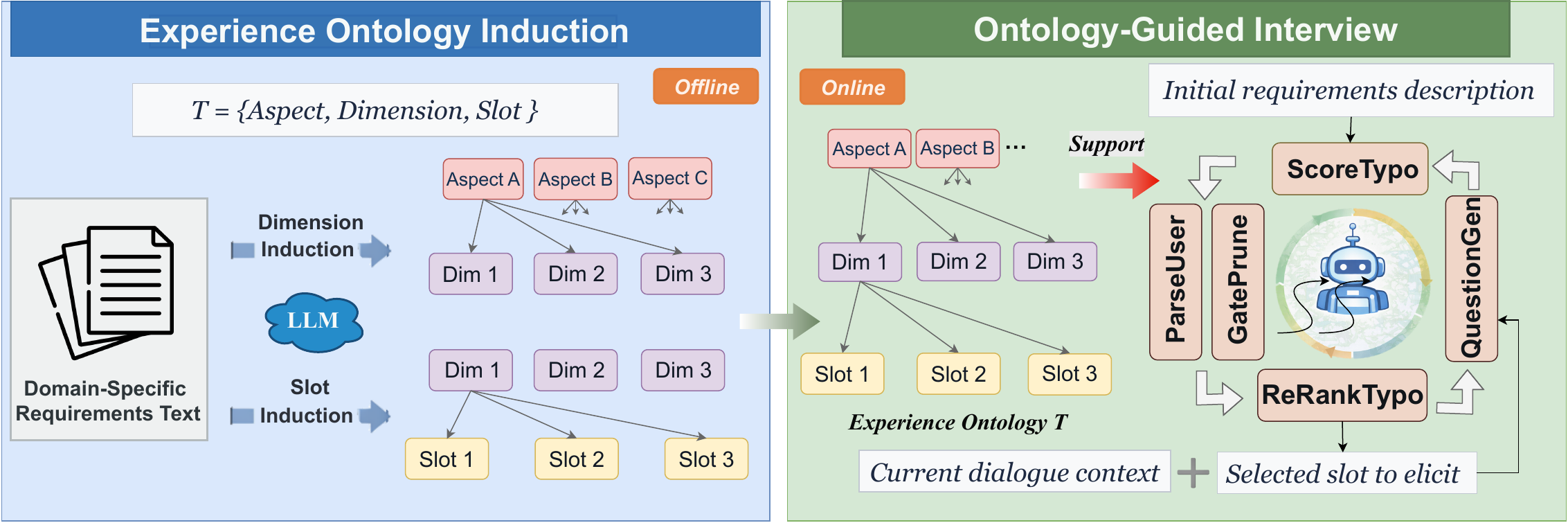}
    \caption{Overview of \ours{} framework.}
    \label{fig:overview}
\end{figure*}

The design principle of \ours{} is explicitly modeling structured interview experience as an ontology and integrating it with LLM-based question generation. Thus, \ours{} consists of two interrelated stages: experience ontology induction and ontology-guide interview. The two stages work in a pipeline manner as shown in Figure~\ref{fig:overview}.

\begin{itemize}
    \item \textbf{Experience Ontology Induction.} Given domain-specific requirement texts $R$, \ours{} automatically analyzes them to induce a hierarchical tree to represent the experience ontology $\mathcal{T}$, which captures recurring requirement aspects, dimensions, and fine-grained clarification slots.
    \item \textbf{Ontology-Guide Interview Process.} Given the induced ontology $\mathcal{T}$, \ours{} conducts a multi-turn dialogue with the stakeholder until no eligible slots remain in $\mathcal{T}$ or a predefined maximum number of turns is reached. At each turn, \ours{} performs four key operations (\ie ParseUser, ScoreOnto, ReRankOnto, and GatePrune) to dynamically select the most relevant yet insufficiently explored requirement slot $S$. It then combines the current dialogue context $C$ with the selected slot $S$ to generate an elicitation question $Q$ for follow-up interaction.
\end{itemize}

\subsection{Experience Ontology Induction}

This stage aims to automatically construct an experience ontology from domain-specific requirement texts. To achieve this, we first define the schema of the experience ontology. Then we design an automated construction pipeline.


\textbf{Ontology Schema.}
\ours{} formulates requirements elicitation experience ontology into a three-layer tree structure. The first layer represents the requirement aspect, which is the highest-level abstraction of system requirements and provides a macro-level partition of the requirement space. 
For example, this layer for the domain of web applications includes \textit{interaction}, \textit{content}, and \textit{style}. 
To ensure the stability of the ontology, the aspect layer is provided by domain experts.
The second layer represents the requirements dimension, which captures the core functional points and serves as a coherent category of requirements. For example, this layer for the web application domain may include \textit{login}, \textit{search}, and \textit{display}. The third layer represents the requirements slot, which corresponds to a clarifiable requirement detail under each dimension. For example, under the \textit{search} dimension, slots may include \textit{filtering options} and \textit{sorting rules}. Formally, the experience ontology is a three-level tree 
$\mathcal{T} = (\mathcal{A}, \mathcal{D}, \mathcal{S}, \pi_D, \pi_S)$, where 
$\mathcal{A}$, $\mathcal{D}$, and $\mathcal{S}$ denote the set of aspects, dimensions, and slots, $\pi_D$ is a mapping that assigns each dimension to one aspect, $\pi_S$ is a mapping that assigns each slot to one dimension.


\textbf{Dimension Induction.} 
The goal of dimension induction is to automatically construct a set of requirement dimensions from domain-specific requirement texts under predefined requirement aspects. Considering that traditional clustering methods suffer from unstable granularity and limited interpretability, \ours{} adopts a gradual expansion approach and employs an LLM to automate this process. Specifically, \ours{} first initializes the ontology using manually defined requirement aspects, with no dimensions assigned under each aspect. Given a requirement description, \ours{} leverages an LLM with a carefully designed prompt $P_d$ to extract its underlying functional dimensions and associate them with the corresponding requirement aspect. The LLM is instructed to either (1) merge the extracted dimensions into an existing dimension when semantic overlap is detected, or (2) introduce a new dimension only when no suitable abstraction exists. Importantly, \ours{} adopts a conservative expansion strategy that prioritizes semantic merging over introducing new dimensions. This strategy prevents uncontrolled growth of the tree and maintains consistent granularity across dimensions.

\begin{Prompt}{User Prompt $P_d$ for Dimension Induction}
Current Ontology Tree:
\{{ontology}\}

New Requirements Text:
\{{requirements}\}

Task:
1) Extract requirement points from the requirements text into the aspect level. 

2) Integrate them into the current tree using the policy (merge > expand > add). Prefer merging into existing

3) Output the strict JSON specified in the system instructions.

\end{Prompt}

\textbf{Slot Induction.} 
Slot induction aims to construct fine-grained and directly clarifiable requirement details under each induced dimension. Given the current two-layer ontology tree (\ie aspect–dimension), \ours{} further employs an LLM guided by a carefully designed prompt $P_s$ to extract concrete requirement slots from domain requirement texts and incrementally associate them with their corresponding dimensions. Formally, a slot is defined as a tuple $S = \langle k, q \rangle$, where $k$ denotes a normalized clarifiable requirement item, and $q$ represents a candidate clarification question associated with that key. During slot induction, semantic conflicts may arise when multiple extracted keys refer to overlapping or highly similar requirement semantics. To resolve such conflicts, \ours{} applies a heuristic disambiguation rule: when multiple keys exhibit semantic overlap, we retain the shorter and more general formulation. This strategy enhances reusability and maintains consistent granularity. \ours{} further regulates the linguistic form of clarification questions to align with different levels of specification. Specifically, when a dimension is merely mentioned without sufficient detail in an analyzed requirements text, clarification questions adopt a binary confirmation form (\eg ``Do you need ...?''). When partial constraints have already been provided, questions shift to an open-ended refinement form (\eg ``What ...?''). This controlled differentiation improves naturalness and avoids premature over-specification during interviews. 

\begin{Prompt}{User Prompt $P_s$ for Slot Induction}
Two-level Ontology:
\{{\textit{current ontology}}\}

New Requirements Text:
\{{\textit{instruction}}\}

Task:
1) Identify which dimensions the requirements text touches.

2) If the text describes in detail, please output a clarifying question; If only mentions, please output "Do you need [X]?" question.

3) Return the strict JSON specified in the system instructions.

4) Omit topics the instruction does not mention.

\end{Prompt}

\subsection{Ontology-Guide Interview}

To integrate the experience ontology with LLM-based dialogue generation, 
\ours{} formulates the interview process as an ontology-constrained iterative decision loop. For each turn in the multi-turn dialogue process, \ours{} first performs four operations 
(\ie ParseUser, ScoreOnto, ReRankOnto, and GatePrune) over the experience ontology to determine the most appropriate requirement slot to be clarified. Subsequently, a QuestionGen operation generates the follow-up elicitation question by combining the selected slot with the current dialogue context. Algorithm~\ref{alg:ontoagent} summarizes the overall process of the ontology-guide interview. The five core operations are detailed below\footnote{All operations are implemented through prompting with rule-based constraints. Detailed prompts are provided in our replication package.}.

\begin{algorithm}[t]
\caption{Ontology-Guided Interview Process}
\label{alg:ontoagent}
\small
\begin{algorithmic}[1]
\State \textbf{Input:} Initial description $u_0$, experience ontology $\mathcal{T}$, max turns $T$, aspect pruning threshold $N$
\State \textbf{Output:} Elicited requirement set $R$

\State $\mathcal{H} \leftarrow \{u_0\}$ 
\State $R \leftarrow \emptyset$
\State $aspect\_no\_need\_count \leftarrow 0$
\State $t \leftarrow 0$
\State Mark all slots in $\mathcal{T}$ as \textsc{Unexplored}
\State $\mathcal{T} \leftarrow \textsc{ScoreOnto}(u_0, \mathcal{T})$  \Comment{\textbf{ScoreOnto}}

\While{$t < T$}
    \Comment{\textbf{Aspect GatePrune}}
    \If{$aspect\_no\_need\_count \geq N$}
        \State $\mathcal{Q} \leftarrow \textsc{CheckAspectPruning}(current\_aspect)$
        \State $\mathcal{A} \leftarrow$ await stakeholder's response
        \State $\mathcal{H} \leftarrow \mathcal{H} \cup \{\mathcal{Q}, \mathcal{A}\}$
        \State $is\_pruned \leftarrow \textsc{ParseUser}(\mathcal{A})$
        \If{$is\_pruned$}
            \State $\mathcal{T} \leftarrow \textsc{AspectPruning}(current\_aspect, \mathcal{T})$
        \EndIf
        \State $aspect\_no\_need\_count \leftarrow 0$
    \EndIf

    \State $\mathcal{S} \leftarrow \textsc{ReRankOnto}(\mathcal{H}, \mathcal{T})$ \Comment{\textbf{ReRankOnto}}

    \If{$\mathcal{S} = \varnothing$}
        \State \textbf{break}
    \EndIf

    \State $\mathcal{Q} \leftarrow \textsc{QuestionGen}(\mathcal{H}, \mathcal{S})$ \Comment{\textbf{QuestionGen}}
    \State $\mathcal{A} \leftarrow$ await stakeholder's response
    \State $\mathcal{H} \leftarrow \mathcal{H} \cup \{\mathcal{Q}, \mathcal{A}\}$

    \State $is\_needed \leftarrow \textsc{ParseUser}(\mathcal{A})$ \Comment{\textbf{Dimension GatePrune}}

    \If{$is\_needed$}
        \State Mark $\mathcal{S}$ as \textsc{Confirmed}
        \State $R \leftarrow R \cup \{\mathcal{S}\}$
    \Else
        \State Mark $\mathcal{S}$ as \textsc{Rejected}
        \State $aspect\_no\_need\_count \leftarrow aspect\_no\_need\_count + 1$
        \State $\mathcal{T} \leftarrow \textsc{ApplyDimensionAndSlotPruning}(\mathcal{S}, \mathcal{T})$
    \EndIf
    \State $t \leftarrow t + 1$
\EndWhile

\end{algorithmic}
\end{algorithm}

\textbf{ParseUser}: This operation interprets stakeholder responses to support pruning and confirmation decisions. It consists of two variants. At the aspect level, it analyzes responses to macro-level confirmation questions 
(\eg whether additional concerns remain under a specific aspect), which enables aspect-level GatePrune. At the dimension level, it determines whether a specific requirement dimension is explicitly confirmed or rejected (\eg \textit{Login is not required}), which supports dimension-level GatePrune.

\textbf{ScoreOnto.} Before the dialogue begins, ScoreOnto assigns initial priority scores 
to nodes in the ontology based on the initial requirement description. This is necessary because early dialogue lacks contextual feedback. 
If the agent simply follows a fixed tree traversal order, the questions in the early stage may deviate from the stakeholder’s primary concerns. By estimating semantic relevance between the initial description and the ontology nodes, ScoreOnto determines an exploration order, ensuring that the interview begins from branches most likely related to the user’s intent.

\textbf{ReRankOnto.} 
During the interview, user responses continuously introduce new context. Therefore, prior priority scores may become outdated. 
Before each turn, ReRankOnto dynamically re-evaluates and reorders candidate slots under the currently active branch based on the updated dialogue history. This operation guarantees that the agent always selects the slot most likely to uncover implicit requirements.


\textbf{GatePrune.} To reduce redundant questioning and accelerate convergence, \ours{} introduces two gate-controlled pruning mechanisms. Aspect-level GatePrune is triggered when no new requirements are elicited after $N$ consecutive inquiries under the same aspect. In this case, \ours{} issues a macro-level confirmation question 
(\eg ``Are there any other requirements related to interaction?''). If the stakeholder explicitly indicates that there are no further concerns, the entire aspect branch is pruned. Dimension-level GatePrune operates at a finer granularity. If the stakeholder explicitly rejects a requirement dimension (\eg stating that registration/login is unnecessary), the corresponding slots in this dimension are immediately pruned. These pruning strategies progressively shrink the search space 
while preserving systematic coverage.

\textbf{QuestionGen.} 
Given the selected slot and the current dialogue history, QuestionGen synthesizes a natural and context-aware interview question. By grounding question generation in structured ontology, \ours{} avoids ad-hoc questioning and maintains interpretability.

\section{Study Design}
To evaluate the performance of the \ours{} framework, we conduct a multi-aspect study to answer five research questions (RQs). This section describes the details of our study, including research questions, baselines, datasets, and metrics.

\subsection{Research Questions}

\textbf{RQ1 (Effectiveness): How effective is \ours{} in conducting requirements elicitation interviews compared with existing baselines?} This RQ aims to evaluate the effectiveness of our proposed framework, \ie improving implicit requirement elicitation 
and questioning efficiency. In experiments, we select GPT-5.1 as the base LLM for our framework. We sample 5 scenarios for each application type from the WenGen-Bench to construct an ontology. The generalization capability across
different LLMs and scenarios number will be further examined in RQ3 and RQ5.

\textbf{RQ2 (Ablation Study): What is the contribution of each component in \ours{}?} This RQ examines how each module contributes to elicitation coverage and questioning efficiency. \ours{} consists of four core components that enhance requirements elicitation interview performance: (1) Experience Ontology, (2) ScoreOnto, (3) ReRankOnto, and (4) GatedPrune\footnote{ParseUser and QuestionGen are not ablated as they are necessary within the pipeline.}. Based on the
experimental setup in RQ1, we conduct ablation studies by incrementally adding these components to a GPT-5.1 base model to analyze their individual contributions. 

\textbf{RQ3 (Sensitivity): How robust is \ours{} across different large language models?} This RQ focuses on the
robustness and generalization ability of \ours{} with respect
to different underlying LLMs. We evaluate whether \ours{} remains effective when substituting GPT-5.1 with other LLMs, validating the model-agnostic design of our framework. Specifically, we follow the prior work~\cite{jin2026reqelicitgym} and select six other LLMs as shown in Table~\ref{tab:llms}.

\textbf{RQ4 (Type-wise Analysis): How does \ours{} perform in eliciting different types of implicit requirements?} 
Implicit requirements may belong to different aspects. Each type may exhibit distinct elicitation characteristics and difficulty levels. This RQ investigates whether \ours{} provides balanced coverage across different requirement types. Specifically, we conduct aspect-level analysis to examine whether \ours{} improves elicitation effectiveness uniformly or exhibits biases toward specific aspects.

\textbf{RQ5 (Scalability): How does the size of induction data affect the performance of \ours{}?} 
Since the experience ontology is constructed from domain requirements text, its quality may depend on the amount of available induction data. This RQ studies the sensitivity of \ours{} to the scale of requirements instructions used for ontology induction. By varying the proportion of the full training data used, we evaluate the scalability and data efficiency of the proposed framework.

\begin{table}[t]
\centering
\small
\caption{The Selected LLMs for RQ3. All models were accessed via official APIs.}
\label{tab:llms}
\begin{tabular}{lllll}
\toprule
\textbf{LLM} & \textbf{Type} & \textbf{Creator} & \textbf{Release} & \textbf{Used} \\
\midrule
Claude Opus 4.5 & Closed & Anthropic & 2025.11 & 2026.02 \\
Gemini 3 Flash & Closed & Google & 2025.12 & 2026.02 \\
DeepSeek V3.2 & Open & DeepSeek & 2025.12 & 2026.02 \\
Kimi K2.5 & Open & Moonshot AI & 2026.01 & 2026.02 \\
GLM-4.7 & Closed & Zhipu AI & 2025.11 & 2026.02 \\
Qwen3 235B & Open & Alibaba & 2025.07 & 2026.02 \\
\bottomrule
\end{tabular}
\end{table}

\subsection{Baselines}
We compare \ours{} against five representative LLM-based elicitation baselines from three recent studies~\cite{KornGV25}~\cite{shen2025requirements}~\cite{jin2026reqelicitgym}. These five baselines all use free-form elicitation approach. Their introductions are provided as follows.

\textbf{Non-CoT.} It follows the standard inference setting adopted in the ReqElicitGym~\cite{jin2026reqelicitgym}. Specifically, an LLM conducts multi-turn dialogue without explicit reasoning instructions. At each turn, the LLM directly generates a clarification or probing question.

\textbf{CoT.} It incorporates Chain-of-Thought (CoT) prompting adopted in the ReqElicitGym~\cite{jin2026reqelicitgym}. Specifically, an LLM is instructed to explicitly reason about the dialogue context before generating each question.

\textbf{LLMREI-short.} It is derived from the zero-shot prompting strategy proposed in LLMREI~\cite{KornGV25}. Specifically, it employs a concise system prompt that instructs the LLM to behave as a professional interviewer and ask one focused question. 

\textbf{LLMREI-long.} It corresponds to the least-to-most prompting strategy proposed in LLMREI~\cite{KornGV25}. Specifically, it adopts a substantially longer and more structured system prompt that embeds explicit interview guidelines, role definitions, and procedural instructions. 

\textbf{Mistake-guided Prompting.} This baseline is derived from the mistake-guided question generation framework~\cite{shen2025requirements}. At each turn, the LLM is provided with the interviewee's utterance together with a predefined interviewer mistake criterion and is instructed to generate a follow-up question. 


\subsection{Datasets}
\textbf{Evaluation Dataset for Interview.} We evaluate \ours{} and all five baselines using the ReqElicitGym~\cite{jin2026reqelicitgym}, an interactive and automatic evaluation environment designed for interviews. ReqElicitGym contains 101 web application scenarios spanning 10 application types. Each scenario consists of an underspecified initial requirement description, a complete final specification, and a set of manually annotated implicit requirements categorized by requirements aspects. In addition, the evaluation environment contains a simulated interactive oracle user and a task evaluator to interact and evaluate an interviewer. Thus, any automated interviewer's approach can be evaluated through interaction with the environment. 

\textbf{Training Dataset for ontology Induction.} To construct the experience ontology, \ours{} requires a corpus of domain-specific requirements descriptions to analyze recurring requirements concerns. In this work, we utilize the train set of WebGen-Bench~\cite{lu2025webgen} as the data source for ontology induction. Notably, all evaluation scenarios in ReqElicitGym are derived from the test set of WebGen-Bench. Therefore, by restricting ontology construction to the WebGen-Bench training data, we ensure strict separation between ontology induction and interview evaluation. This design effectively prevents data leakage and guarantees that the induced ontology reflects generalizable elicitation experience rather than memorization of evaluation scenarios.

\subsection{Evaluation Metrics}\label{subsec:metrics}

We evaluate the performance using two complementary measures: Implicit Requirements Elicitation Ratio (IRE)~\cite{jin2026reqelicitgym} and 
Turn-discounted Key Question Rate (TKQR)~\cite{fangzcs_cognicode_2026}. 
IRE measures overall coverage of implicit requirements, while TKQR evaluates questioning efficiency by emphasizing the ordering. These metrics capture both the effectiveness and efficiency of elicitation interviews.

\textbf{Implicit Requirements Elicitation Ratio (IRE).} Let $\mathcal{R}$ denote the set of ground-truth implicit requirements for a scenario, and let $\hat{\mathcal{R}}_{\le T}$ denote the set of implicit requirements elicited by the interviewer when the interaction ends at turn $T$. The metric is computed as:
\begin{equation}
\mathrm{IRE} = \frac{|\hat{\mathcal{R}}_{\le T}|}{|\mathcal{R}|}.
\end{equation}

To enable fine-grained analysis, aspect-level IRE is computed by restricting $\mathcal{R}$ and $\hat{\mathcal{R}}_{\le T}$ to implicit requirements within each requirement aspect. This aspect-wise breakdown facilitates examining whether an elicitation approach achieves balanced coverage across different requirement aspects.

\textbf{Turn-discounted Key Question Rate (TKQR).} TKQR evaluates questioning efficiency by rewarding early elicitation of key implicit requirements and penalizing delayed or redundant questioning. Let $n$ denote the number of dialogue turns before the interviewer stops asking questions, and let $K = |\mathcal{R}|$ denote the total number of annotated implicit requirements for the scenario. A hit indicator sequence $H=(h_1,\dots,h_n)$ is constructed, where $h_i=1$ if the interviewer elicits a previously unelicited implicit requirement at turn $i$, and $h_i=0$ otherwise. 
The discounted cumulative gain is computed as:
\begin{equation}
\mathrm{DCG}_n = \sum_{i=1}^{n} \frac{h_i}{\log_2(i+1)}.
\end{equation}
To normalize across scenarios with different numbers of implicit requirements, the ideal discounted cumulative gain is computed as:
\begin{equation}
\mathrm{IDCG}_n = \sum_{i=1}^{\min(n,K)} \frac{1}{\log_2(i+1)}.
\end{equation}

TKQR is then calculated as:
\begin{equation}
\mathrm{TKQR} = \frac{\mathrm{DCG}_n}{\mathrm{IDCG}_n}.
\end{equation}

TKQR ranges in $[0,1]$, where higher values indicate earlier prioritization of key elicitation questions during the interaction.
\section{Results and Analysis}~\label{sec:result}

\textbf{RQ1 (Effectiveness): How effective is \ours{} in conducting requirements elicitation interviews compared with existing baselines?} 

\textbf{Setup.} We evaluate five baselines and our \ours{} on 101 website requirements elicitation scenarios in ReqElicitGym. The evaluation metrics are described in Section~\ref{subsec:metrics}, \ie IRE and TKQR. For all metrics, higher scores represent better performance.

\textbf{Results.} Table~\ref{tab:rq1} shows the experimental results on ReqElicitGym. 

\textbf{Analyses.} \textbf{(1) \ours{} improves implicit requirement elicitation coverage.} Table~\ref{tab:rq1} shows that \ours{} achieves an IRE of 0.69, outperforming all baselines.
The strongest baseline (\ie Mistake-guided Prompting) reaches 0.52. Compared with it, \ours{} gain a relative improvement by 33\%, demonstrating its ability to systematically uncover missing implicit requirements. The performance gap between free-form baselines and \ours{} indicates that relying solely on LLMs to chat is insufficient for structured interviews. Instead, it is essential to explicitly model interview experience and integrate it with LLMs for achieving higher elicitation coverage.
\textbf{(2) \ours{} also achieves the highest questioning efficiency.}
In terms of TKQR, \ours{} achieves 0.59, outperforming all baselines. It gains a 21\% relative improvement over the best baseline (\ie LLMREI-short). This improvement indicates that \ours{} not only elicits more implicit requirements but also identifies them earlier in the interaction.
Since TKQR penalizes delayed discovery, the higher score demonstrates that \ours{} can elicit high-value requirement dimensions in early dialogue turns.


\begin{boxK}
\small \faIcon{pencil-alt} \textbf{Answer to RQ1:} \ours{} significantly outperforms the baselines in both implicit requirement elicitation effectiveness and questioning efficiency. In particular, it gains a relative improvement by 33\% in IRE and 21\% in TKQR. The significant improvements prove our ontology-enhanced approach is more promising.
\end{boxK}

\begin{table}[t]
    \centering
    \caption{Overall effectiveness comparison on ReqElicitGym. Relative improvement is computed against the best-performing baseline.}
    \begin{tabular}{lcc}
    \toprule
    \textbf{Approach} & \textbf{IRE} & \textbf{TKQR} \\
    \midrule
    Non-CoT & 0.13 & 0.09 \\
    CoT & 0.08 & 0.19 \\
    LLMREI-short (RE'25) & 0.39 & 0.49 \\
    LLMREI-long (RE'25) & 0.38 & 0.09 \\
    Mistake-guided Prompt (RE'25) & 0.52 & 0.48 \\
    \midrule
    \textbf{\ours{} (Ours)} & \textbf{0.69 ($\uparrow$ 33\%)}& \textbf{0.59 ($\uparrow$ 21\%)}\\
    \bottomrule
    \end{tabular}
    \label{tab:rq1}
\end{table}

\textbf{RQ2 (Ablation Study): What is the contribution of each component in \ours{}?} 

\textbf{Setup.} Starting from a base LLM (\ie GPT-5.1), we progressively add the four core components of \ours{}, \ie Experience Ontology, ScoreOnto, ReRankOnto, and GatedPrune. All variants are evaluated using the same ReqElicitGym environment. We report the same two metrics as in RQ1, \ie IRE and TKQR.

\textbf{Results.} The experimental results of the ablation study are shown in Table~\ref{tab:rq2}.

\textbf{Analyses.} \textbf{(1) Experience Ontology provides the largest performance improvement.} Adding the experience ontology dramatically improves IRE from 0.13 to 0.41 and TKQR from 0.09 to 0.34.
This substantial improvement indicates that explicitly modeling the requirements interview experience to guide the LLM is far more effective than relying solely on free-form generation. The ontology can effectively transform the elicitation process from LLM-based chat into structured interviews. \textbf{(2) ScoreOnto and ReRankOnto further improve early-stage alignment and efficiency.} Introducing ScoreOnto further increases IRE to 0.58 and TKQR to 0.37. These indicate that this component enables the agent to focus on more relevant requirement dimensions at early dialogue turns. Adding ReRankOnto yields additional gains, improving IRE to 0.64 and TKQR to 0.62. Although the improvement is relatively small, it also demonstrates the necessity of dynamically adjusting questioning direction based on the accumulated dialogue context.
\textbf{(3) Gated Pruning improves efficiency while maintaining coverage.} Incorporating Gated Pruning further improves IRE to 0.69 and TKQR to 0.59. The increase in TKQR indicates that pruning reduces redundant questions, enabling earlier elicitation of key requirements. Importantly, the pruning does not harm coverage. Instead, it improves IRE, suggesting that eliminating ineffective questioning paths helps concentrate the exploration on valid dimensions.

\begin{table}[t]
    \centering
    \caption{Ablation study of \ours{} on ReqElicitGym.
    Each row incrementally adds one component to the base LLM.}
    \begin{tabular}{lcc}
    \toprule
    \textbf{Approach} & \textbf{IRE} & \textbf{TKQR} \\
    \midrule
    Base LLM (GPT-5.1) & 0.13 & 0.09 \\
    \quad + Experience Ontology& 0.41 & 0.34\\
    \quad + ScoreOnto& 0.58 & 0.37 \\
    \quad + ReRankOnto& 0.64 & 0.52 \\
    \quad + GatePrune& \textbf{0.69} & \textbf{0.59} \\
    \bottomrule
    \end{tabular}
    \label{tab:rq2}
\end{table}

\begin{boxK}
\small \faIcon{pencil-alt} \textbf{Answer to RQ2:} Four modules are essential for the performance of our approach. Experience Ontology provides the largest gain both in coverage and efficiency. ScoreOnto and Re-ranking improve alignment and efficiency during early-stage interaction. Gated Pruning further enhances efficiency by reducing redundant questions without harming coverage.
\end{boxK}

\textbf{RQ3 (Robustness): How robust is \ours{} across different large language models?} 

\textbf{Setup.} To evaluate the robustness of \ours{} with respect to different backbone LLMs, we substitute GPT-5.1 with several representative LLMs in Table~\ref{tab:llms}. We report IRE and TKQR under the same evaluation environment as in RQ1.

\textbf{Results.} Table~\ref{tab:rq3_projection} summarizes the overall performance of \ours{} under different backbone LLMs. 

\textbf{Analyse.} \textbf{(1) \ours{} maintains competitive performance across diverse backbones.} Across six representative LLMs, \ours{} consistently achieves IRE between 0.55 and 0.68, and TKQR between 0.47 and 0.74. Despite variations in model scale and training paradigms, all backbones demonstrate effective implicit requirement elicitation when integrated with \ours{}. This indicates that the \ours{} generalizes across heterogeneous LLM capabilities rather than relying on a single specific backbone. \textbf{(2) Backbone quality affects performance, but does not solely determine outcomes.} Models with stronger reasoning and planning capabilities (\eg Claude Opus 4.5) tend to achieve higher IRE or TKQR scores. However, the gap between the highest and lowest IRE remains moderate (\ie 0.13), suggesting that backbone differences do not fundamentally alter elicitation effectiveness. This indicates that \ours{} reduces dependence on purely implicit reasoning ability of LLMs. 


\begin{table}[t]
\centering
\caption{Performance of \ours{} on different LLM backbones.}
\begin{tabular}{lcc}
\toprule
\textbf{Backbone LLM} & \textbf{IRE} & \textbf{TKQR} \\
\midrule
Claude Opus 4.5 + \ours{} & 0.63 & 0.74 \\
Gemini 3 Flash + \ours{} & 0.64 & 0.48 \\
DeepSeek V3.2 + \ours{} & 0.55 & 0.71 \\
Kimi K2.5 + \ours{} & 0.64 & 0.59 \\
GLM-4.7 + \ours{} & 0.63 & 0.47 \\
Qwen3 235B + \ours{} & 0.68 & 0.52 \\
\bottomrule
\end{tabular}
\label{tab:rq3_projection}
\end{table}

\begin{boxK}
\small \faIcon{pencil-alt} \textbf{Answer to RQ3:} \ours{} exhibits strong robustness across different backbone LLMs. Although backbone quality leads to moderate variation, overall elicitation effectiveness and efficiency remain stable. This indicates that \ours{} plays the primary role in determining performance rather than specific model capability.
\end{boxK}

\textbf{RQ4 (Requirement-Type-wise Analysis): How does \ours{} perform in eliciting different types of implicit requirements?} 

\textbf{Setup.}
Following the ReqElicitGym~\cite{jin2026reqelicitgym}, the implicit requirements are categorized into three dimensions: \textit{interaction}, \textit{content}, and \textit{style}. For all baselines and \ours{}, we compute aspect-level IRE by restricting the ground-truth and elicited requirement sets to each requirement type. 


\textbf{Results.} Table~\ref{tab:rq4_aspect} presents the aspect-level elicitation performance of five baselines and \ours{}. 

\textbf{Analyse.} \textbf{(1) \ours{} consistently outperforms all baselines across all requirement types.} \ours{} achieves the highest IRE in \textit{interaction} (\ie 0.74), \textit{content} (\ie 0.64), and \textit{style} (\ie 0.55), demonstrating its ability to provide balanced coverage across heterogeneous requirement dimensions. In contrast, baseline methods exhibit uneven performance, particularly struggling with \textit{style} requirements. \textbf{(2) \textit{Style} requirements are particularly challenging for baseline methods.} Both Non-CoT and CoT achieve near-zero performance on \textit{style} requirements ($<0.01$), indicating that free-form conversational generation rarely explores aesthetic or presentation-related aspects without explicit structural guidance.
Even structured prompt-based baselines only achieve 0.05, 0.09, and 0.17, respectively, suggesting that prompt engineering alone is insufficient for eliciting stylistic or non-functional preferences. By contrast, \ours{} improves \textit{style}-related IRE dramatically to 0.55.

\begin{table}[t]
    \centering
    \caption{Aspect-level implicit requirement elicitation performance.
    IRE\_int, IRE\_con, and IRE\_sty denote the elicitation ratios for \textit{interaction}, \textit{content}, and \textit{style} requirements, respectively.}
    \begin{tabular}{lccc}
    \toprule
    \textbf{Methods} & \textbf{IRE\_int} & \textbf{IRE\_con} & \textbf{IRE\_sty} \\
    \midrule
    Non-CoT & 0.19 & 0.13 & $<0.01$ \\
    CoT & 0.12 & 0.09 & $<0.01$ \\
    LLMREI-long & 0.56 & 0.50 & 0.05 \\
    LLMREI-short & 0.59 & 0.51 & 0.09 \\
    Mistake-guided Prompt & 0.70 & 0.65 & 0.17 \\
    \ours{} & \textbf{0.74} & \textbf{0.64} & \textbf{0.55} \\
    \bottomrule
    \end{tabular}
    \label{tab:rq4_aspect}
\end{table}

\begin{boxK}
\small \faIcon{pencil-alt} \textbf{Answer to RQ4:} \ours{} achieves balanced and substantial improvements across all requirement types. The most significant gain is observed for \textit{style} requirements, which are largely neglected by baseline methods.
\end{boxK}

\textbf{RQ5 (Scalability): How does the size of induction data affect the performance of \ours{}?}

\textbf{Setup.} For each application type in the WenGen-Bench training set, we randomly sample 5, 10, 15, and 20 scenarios to induce the ontology, respectively. All induced ontologies are then fixed and evaluated under the same ReqElicitGym environment as in RQ1. 
For each data size setting, we report IRE and TKQR.

\begin{figure}
    \centering
    \includegraphics[width=0.9\linewidth]{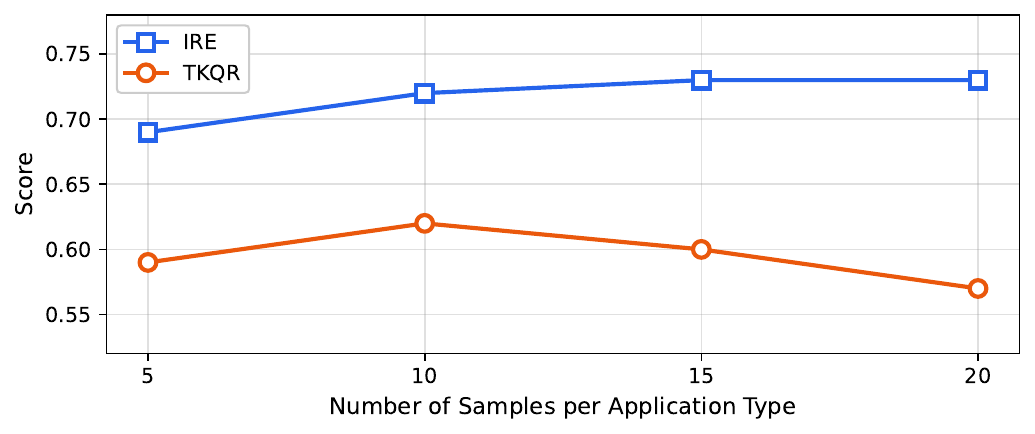}
    \caption{Impact of ontology induction data size on the performance.}
    \label{fig:rq5_scalability}
\end{figure}

\textbf{Results.} Figure~\ref{fig:rq5_scalability} shows the performance of \ours{} under different induction data sizes. 

\textbf{Analyse.} \textbf{(1) Increasing induction data moderately further improves elicitation coverage.} As the number of induction scenarios increases from 5 to 15, IRE improves from 0.69 to 0.73, indicating that richer domain data helps construct a more comprehensive ontology and improves coverage of implicit requirements. \textbf{(2) Excessive induction data may reduce questioning efficiency.} When the data size further increases to 20 scenarios, TKQR decreases from 0.60 to 0.57. This suggests that an overly large ontology expands the search space and introduces more candidate slots, making early-stage question selection less focused. 

\begin{boxK}
\small \faIcon{pencil-alt} \textbf{Answer to RQ5:} Increasing induction data improves elicitation coverage, showing that \ours{} benefits from richer ontology construction. However, overly large induction data may negatively affect questioning efficiency due to an expanded search space.
\end{boxK}

\section{Human Evaluation and Case Analysis}
In this section, we complement our quantitative results in Section~\ref{sec:result} with a human evaluation study and cross-case analysis. We aim to examine the practical effectiveness and behavioral characteristics of \ours{} beyond metrics.

\subsection{Human Evaluation}

The experiments presented in Section~\ref{sec:result} were conducted within the ReqElicitGym simulated environment.
To further examine the practical effectiveness of \ours{}, we conducted a complementary human evaluation study.

\textbf{Participants.}
We recruited 6 participants (4 graduate students and 2 senior undergraduate students) with software engineering backgrounds. All participants had prior coursework or project experience in requirements analysis. None of them were involved in the development of \ours{}.

\textbf{Study Design.} We designed an experiment involving real interviews and adopted a within-subject design~\cite{charness2012experimental}. Specifically, each participant acted as a stakeholder in predefined scenarios and interacted with four approaches (\ie LLMREI-short, LLMREI-long, Mistake-guide Prompting and \ours{}). The predefined scenarios come from a previous study~\cite{ferrari2020sapeer} and have been used in multiple requirements-related studies~\cite{KornGV25}~\cite{ferrari2020sapeer}~\cite{voria2025recover}. They include: \textbf{(1)Salon Scenario.} A hair and nail salon seeking a digital solution for managing appointments and employee scheduling. \textbf{(2) Ski Resort Scenario.} A ski resort requiring a digital booking and business management platform for its three locations. Before each session, participants received a description outlining their role as a stakeholder and instructions on what to do after the interview. Then the chatbot, based on the above approaches, conducted the interview separately. Each participant completed the interview session with them. Each interview session lasted around 20 minutes. Upon completion, participants were asked to fill out a questionnaire evaluating the effectiveness of the approach. Participants rated the effectiveness on a 7-point Likert scale\footnote{1 = strongly disagree, 7 = strongly agree} across three dimensions, \ie Elicitation Effectiveness, Questioning Efficiency, and Questioning Adaptability. 

\textbf{Result and Analysis.} The results of the human evaluation are shown in Table~\ref{tab:human_eval}. Our \ours{} is better than all baselines in three aspects. Specifically, \ours{} outperforms the SOTA baselines by 15\% in elicitation effectiveness, 10\% in questioning efficiency, and 19\% in questioning adaptability. All the p-values are substantially smaller than 0.05, which shows the improvements are statistically significant. The improvements prove the superiority of our \ours{} in assisting requirements elicitation interviews. Besides, we acknowledge the number of participants is relatively limited. The human evaluation is intended to complement the large-scale quantitative results in Section~\ref{sec:result}, and the consistent trends across both evaluations further demonstrate the effectiveness of \ours{}.

\begin{table}[t]
\centering
\caption{Human evaluation results (mean Likert scores, 1–7 scale).}
\begin{tabular}{lccc}
\toprule
Approach & Effectiveness & Efficiency & Adaptability \\
\midrule
LLMREI-Short & 4.58 & 4.92 & 5.14 \\
LLMREI-Long  & 4.47 & 3.26 & 4.74 \\
Mistake-guided Prompt & 5.11 & 5.26 & 5.08 \\
\textbf{\ours{}}   & \textbf{5.87 ($\uparrow$ 15\%)} & \textbf{5.79 ($\uparrow$ 10\%)} & \textbf{6.12 ($\uparrow$ 19\%)} \\
\bottomrule
\end{tabular}
\label{tab:human_eval}
\end{table}

\subsection{Case Analysis}
\begin{figure*}
    \centering
    \includegraphics[width=0.99\linewidth]{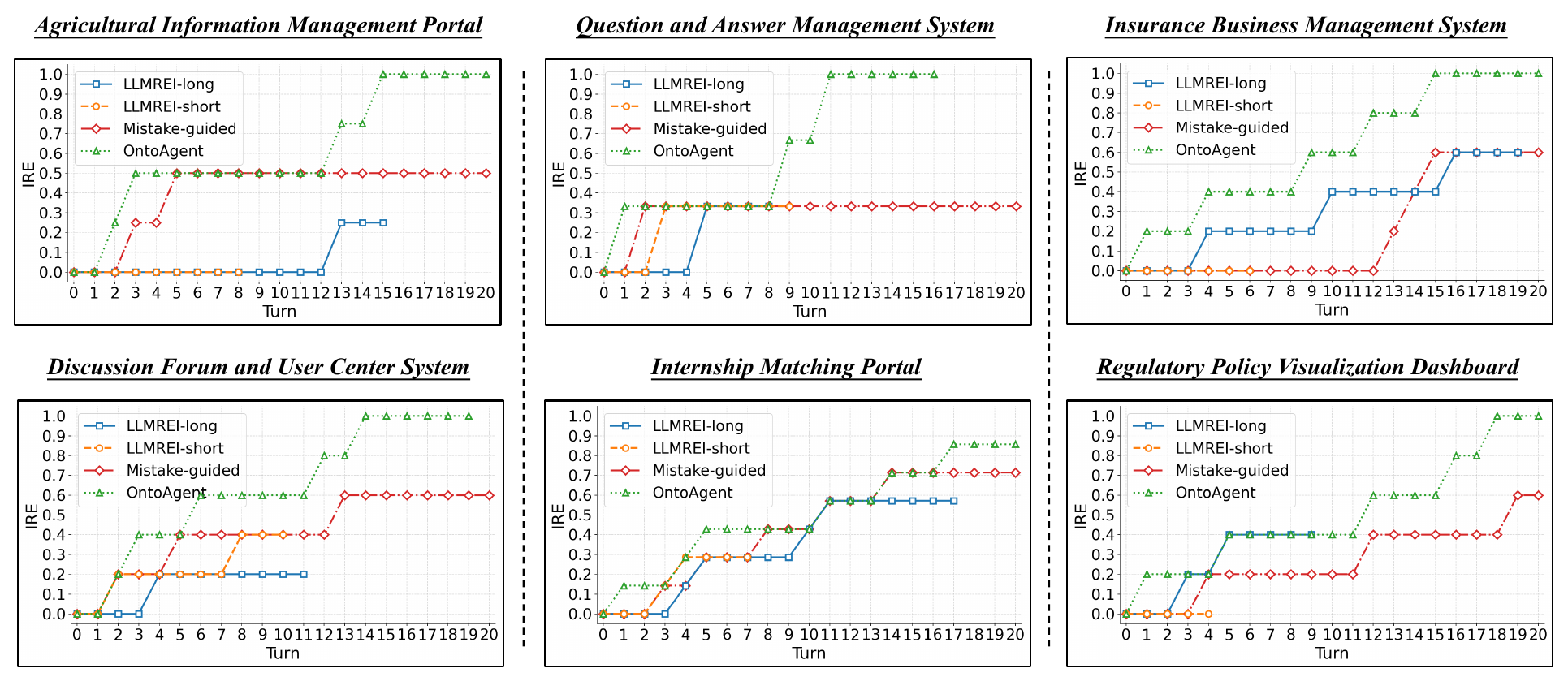}
    \caption{Turn-level IRE progression across six representative scenarios. }
    \label{fig:case_study}
\end{figure*}

To better understand how \ours{} and compared baselines behave, we conduct a cross-case analysis on six representative scenarios in ReqElicitGym. Figure~\ref{fig:case_study} presents the turn-level IRE progression for three baselines and our \ours{}. From the examples, we obtain the following findings. \textbf{(1) Faster Early Convergence.} \ours{} demonstrates substantially faster early-stage IRE growth, often eliciting key implicit requirements within the first 3–5 turns, while free-form baselines typically exhibit prolonged stagnation in the initial dialogue phase. This suggests that \ours{} enables \ours{} to immediately focus on high-impact requirements dimensions rather than spending early turns on surface-level clarification. \textbf{(2) Higher Final Coverage.} \ours{} consistently achieves the highest final IRE, frequently approaching near-complete implicit requirement coverage. In contrast, free-form baselines often plateau at moderate coverage levels, indicating omission of less salient but critical requirements aspects. This highlights the advantage of explicitly modeling the requirement space as a structured, hierarchical ontology to prevent systematic blind spots. \textbf{(3) Reduced Mid-Turn Stagnation.} The progression curves of \ours{} exhibit steady, step-wise improvements with fewer long plateaus, reflecting systematic traversal of requirements dimensions. In comparison, the baselines display ad-hoc jumps and extended stagnation phases, suggesting unsystematic exploration and redundant questioning.

\section{Threats to Validity}

\textbf{Construct Validity.} Construct validity concerns the relationship between treatment and outcome.  Potential threats arise from the metrics adopted and the datasets involved in ontology induction and evaluation. The first threat is that these metrics may not fully capture all qualitative aspects of interview performance. To mitigate this concern, we employ two complementary quantitative metrics (\ie IRE and TKQR), which measure elicitation effectiveness and turn-level efficiency from different perspectives, and further complement them with a human evaluation study that assesses elicitation quality across multiple dimensions using Likert-scale ratings. The second threat relates to the evaluation benchmark and the data used for ontology induction. Although ReqElicitGym is a simulated requirements elicitation environment, it has been validated in prior work~\cite{jin2026reqelicitgym} as a reliable benchmark for assessing conversational requirements interview competence. The data used for ontology construction may also influence the performance of \ours{}. To address this concern, we enforce strict separation between ontology induction data and interview evaluation data, ensuring that the induced ontology is constructed without exposure to any evaluation instances. In addition, we conduct an empirical study to examine the impact of induction data size on the performance of \ours{}, which further strengthens confidence that the observed improvements are not attributable to data leakage or memorization.

\textbf{Internal Validity.} Internal validity addresses potential threats to the way the study was conducted. First, model configuration and evaluation settings may influence performance. To ensure stability and reproducibility, all LLMs are evaluated using greedy decoding. In addition, the maximum number of dialogue turns is uniformly set to 20 across all approaches to guarantee fair comparison under identical interaction budgets. Second, the re-implementation of baseline methods may introduce bias. To mitigate this threat, we strictly follow the original prompt designs and experimental protocols described in the corresponding papers without modification. No additional tuning or optimization is applied to favor \ours{}. Third, the human evaluation process may be affected by participant bias or learning effects when interacting with multiple systems sequentially. To reduce this influence, we adopt a within-subject design and ensure that all participants read the scenario descriptions beforehand. We acknowledge that residual subjective bias may still exist. However, combining quantitative metrics with human evaluation helps reduce the impact of any single evaluation source.

\textbf{External Validity.}
External validity considers the generalizability of our findings. A primary threat is that our evaluation relies on ReqElicitGym as the experimental environment. Although ReqElicitGym covers diverse application types, it is still confined to the web application domain. To mitigate this concern, we complement the benchmark-based evaluation with a human evaluation study conducted on two additional widely used real-world scenarios outside the ReqElicitGym setting. This helps validate the effectiveness of \ours{} under more natural and varied interaction contexts. 


\section{Conclusion}

We propose \ours{}, an ontology-enhanced requirements elicitation agent that transforms free-form LLM interviews into structured and interpretable inquiry processes. By explicitly modeling interview experience as a hierarchical ontology and guiding question selection through dynamic prioritization and pruning, \ours{} enables systematic exploration of implicit requirements. Experiments on 101 scenarios demonstrate significant improvements over strong LLM baselines in both elicitation coverage and questioning efficiency. Human evaluation further validates its practical effectiveness. \ours{} also demonstrate the importance of ontology-guided LLM agents for advancing automated requirements development.

\section{Data Availability}
Our source code, dataset, and a lightweight tool are available at \url{https://anonymous.4open.science/r/TypoAgent-RE2026}.

\useunder{\uline}{\ul}{}

\normalem
\balance
\bibliographystyle{IEEEtran}
\bibliography{main}

\end{document}